\documentclass[10pt]{article}
\usepackage{amssymb}
\usepackage{amsmath}
\usepackage{enumerate}
\usepackage{multirow} 
\usepackage{graphicx,float,lipsum}
\usepackage{wrapfig}
\usepackage{multicol}
\usepackage{braket}
\usepackage[left=2cm,right=1.5cm,top=3cm,bottom=3cm]{geometry}
\usepackage{xcolor}

\title{\bf Renormalization Group Approach to the Continuum Limit of Matrix Models of Quantum Gravity with Preferred Foliation}
\author{{\bf Alicia Castro}$^1$ and {\bf Tim A. Koslowski}$^2$
\\
{\it Email:} $^1$ alicia.castro@science.ru.nl, \\
IMAPP,
Radboud University,\\
Heyendaalseweg 135, 6525 AJ Nijmegen, The Netherlands\\
{\it Email:} $^2$ t.a.koslowski@gmail.com\\
Universit\"at W\"urzburg, Emil-Hilb-Weg 22\\
97074 W\"urzburg, Germany
}
\date{\vspace{-5ex}}

\begin{document}

\maketitle

\begin{abstract}
\noindent This contribution is not intended as a review but, by suggestion of the editors, as a glimpse ahead into the realm of dually weighted tensor models for quantum gravity. This class of models allows one to consider a wider class of quantum gravity models, in particular one can formulate state sum models of spacetime with an intrinsic notion of foliation. The simplest one of these models is the one proposed by Benedetti and Henson \cite{Benedetti:2008}, which is a matrix model formulation of two-dimensional Causal Dynamical Triangulations (CDT). In this paper we apply the Functional Renormalization Group Equation (FRGE) to  the Benedetti-Henson model with the purpose of investigating the possible continuum limits of this class of models. Possible continuum limits appear in this FRGE approach as fixed points of the renormalization group flow where the size of the matrix acts as the renormalization scale. Considering very small truncations, we find fixed points that are compatible with analytically known results for CDT in two dimensions. By studying the scheme dependence of our results we find that precision results require larger truncations than the ones considered in the present work. We conclude that our work suggests that the FRGE is a useful exploratory tool for dually weighted matrix models. We thus expect that the FRGE will be a useful exploratory tool for the investigation of dually weighted tensor models for CDT in higher dimensions. 
\end{abstract}
\medskip

\begin{multicols}{2}

\section{Introduction}
    
The construction of a unified theory that contains the two most successful branches of modern physics, i.e. General Relativity (GR) and Quantum Field Theory (QFT) in a curved spacetime, as appropriate limits has been ongoing for more than eighty years and sparked many approaches to the so-called problem of quantum gravity. A complete list of these approaches goes beyond the scope of this introduction. The approaches range from the very conservative application of QFT \cite{Hawking:1979ig, Donoghue:1994dn} methods to theories of gravity and the asymptotic safety conjecture \cite{Reuter:1996cp,Reuter:2012id,Eichhorn:2018yfc} over refined applications of quantization rules, such as loop quantum gravity \cite{Thiemann:2007zz}, spin foams \cite{Perez:2012wv}, group field theories \cite{Freidel:2005qe} and tensor models \cite{Rivasseau:2012yp,Rivasseau:2013uca,Rivasseau:2016zco,Rivasseau:2011hm} to significantly less conservative approaches like emergent gravity \cite{Barcelo:2005fc}, holographic duality \cite{Witten:1998qj} and to searches for so-called theories of everything such as string theory \cite{Polchinski:1998rq}. None of these approaches has produced a completely satisfactory answer to the problem of quantum gravity as of now. However, most approaches possess some built-in features but all known approaches come with intrinsic short-comings that have to be overcome before qualifying the particular approach as a candidate theory of quantum gravity. This suggests to combine approaches with different built-in strengths with the goal of obtaining a new approach that combines the best of both.

The present contribution intends precisely this by combining the systematic renormalization group investigation of tensor models for quantum gravity with the success of CDT in producing phases in which the partition function is dominated by extended geometries. This combination is most straightforwardly possible when CDT is formulated as a dually weighted tensor model. Before going into detail, let us take a step back and describe the big picture schematically:

Tensor models of quantum gravity are based on the idea of Euclidean lattice quantum gravity, i.e. a partition function approach in which one sums over Boltzmann factors for spacetimes that are constructed from discrete building blocks. The continuum limit of these partition functions is taken as the limit in which the size of the building blocks is sent to zero while the total volume of the spacetime is held fixed. This implies that the number of building blocks has to diverge when taking the continuum limit, thus indicating that one needs to consider the renormalization group flow of these partition functions in order to study possible phase transitions. A particularly useful tool for the systematic investigation of non-perturbative renormalization group flow is the FRGE, which takes the form of a simple one-loop equation that describes an interpolation between a bare action and the quantum effective action \cite{Wetterich:1992}. To apply this powerful tool to Euclidean lattice gravity it is very useful to exploit the duality between the Feynman-graphs of (un)-colored tensor models and discrete geometries. This duality allows one to identify the Feynman amplitude of the (un)-colored tensor model with the Boltzmann factor of the associated discrete gravity partition function and hence allows a translation from the tensor model action to the discrete gravity action, which takes the form of a Regge action \cite{Regge:1961px}. Therefore the conjecture that the large $N$-limit of the tensor model translates into the continuum limit of the discrete gravity partition function. Hence, the investigation of continuum limits of lattice quantum gravity is translated into the investigation of the possible large $N$-limits of tensor models, which can be investigated systematically using the FRGE. 

This rigorous connection between the continuum limit of Euclidean lattice quantum gravity and the large $N$-renormalization group flow of tensor model actions is an invaluable intrinsic feature of the tensor model approach to quantum gravity; and the systematic investigation of these continuum limits with the FRGE is particularly convenient. Unfortunately, the extended geometries that are approximated in the continuum limits that have been investigated so far possess dimension two or less. In other words, so far no state sum model of discrete geometry is known to coarse grain to a model of extended spacetime geometry in more than two dimensions.

There are however numerical indications that $d$-dimensional CDT and its counter part Euclidean Dynamical Triangulations (EDT) do coarse grain to extended higher dimensional geometries (for $2\leq d \leq 4$). This can be heuristically understood as the fact that the foliation in CDT and the volume term in EDT implement additional terms in the Boltzmann factor for discrete geometry which change the universality class of the model. Moreover, there exist tensor model formulations of CDT and EDT in the literature. The novelty in these models is that they possess a nontrivial propagator, which implements a dual weighting of the Feynman graphs of these tensor models. Hence, one can use the FRGE to investigate the continuum limits of CDT and EDT by studying the renormalization group flow of tensor models with dual weights. This is the motivation for the work presented in the present contribution.

As a first step, we consider a dually weighted matrix model proposed by Benedetti and Henson whose partition function is dual to two-dimensional CDT \cite{Benedetti:2011nn}. By doing this we follow a strategy that was used when first applying the FRGE to tensor models \cite{Eichhorn:2013isa}, where matrix models for two-dimensional Euclidean quantum gravity were considered to introduce the setup, develop the technique and to compare with the analytic results known from constructive approaches to two-dimensional Euclidean quantum gravity, which serve as a bench mark. This allows us to test a setup (the systematic FRGE investigation to dually weighted tensor models) that is readily available in higher dimensions, in particular in 3+1 dimensions \cite{Benedetti:2011nn}, but at the same time is understood analytically, providing benchmark results for the FRGE calculation which we can use to gauge this setup.

This contribution is organized as follows: In the following section (section \ref{sec:preliminaries}) we provide the necessary background on dually weighted tensor models, the particular model proposed by Benedetti and Henson and the foundations of the application of the FRGE to tensor models. We provide the derivation of the beta functions in section \ref{sec:betaFunctions}. We perform a fixed point analysis and study of scheme dependence in section \ref{sec:fixedPointAnalysis}. We summarize our results in section \ref{sec:conclusions} and briefly discuss their implications for future investigations on dually weighted tensor models for quantum gravity.

 


\section{Preliminaries}\label{sec:preliminaries}

Random tensor models are by now an established approach to Euclidean quantum gravity. However, to fully appreciate the way in which dually weighted matrix models provide an approach to quantum gravity with a preferred time slicing one needs to take a step back and consider the foundations of random tensor models.

\subsection{Tensor Models and Dual Weights}
The random tensor model approach to quantum gravity is based on the basic observation that the Feynman graphs of so-called uncolored random tensor models possess a geometric interpretation in terms of tessellations of piece-wise linear pseudo-manifolds, as do some so-called colored tensor models. The uncolored models are defined for tensors $T_{i_1i_2,...i_k}$ and their complex conjugates $\bar{T}_{i_1i_2i_3...i_k}$ through the symmetry of the action $S[T,\bar T]$ under the $U^k(N)$ transformations
\begin{equation}
    T_{i_1i_2...i_k}\mapsto {(U_1)}_{i_1}^{j_1}{(U_2)}_{i_2}^{j_2}...{(U_k)}_{i_k}^{j_k} T_{j_1j_2...j_k}.
\end{equation}
This symmetry implies that the action can be expanded in terms of generalized trace invariants in which the first index $i_1$ of each tensor $T$ must be contracted with the first index of a complex conjugated tensor $\bar T$, and similarly the second index $i_2$ and all further indices $i_l$. These generalized traces can be represented as colored graphs where each tensor $T$ is represented by a white vertex and each complex conjugate tensor $\bar T$ is represented by a black vertex and each contraction of vertices by an index $i_l$ is represented by an edge of color $l$ connecting the vertices associated with the two contracted tensors. Such colored graphs are then dual to piecewise linear pseudo-manifolds: Each vertex is associated with a $(k-1)$-simplex and the adjacent edges are associated with a gluing of the colored $(k-2)$-simplices in the boundary of the two $(k-1)$ simplices. Moreover, closed two-colored sub-graphs are associated with the gluing of $(k-3)$-simplices in the boundary of the associated $(k-2)$-simplices. Analogously, closed three- and more-colored subgraphs are associated with the gluing of simplices of even lower dimension. The generalized trace-invariants of a rank $k$ tensor model can thus be interpreted as tessellations of piecewise linear $(k-1)$-dimensional manifolds. 

We can now perform the analogous identification for the Feynman graphs generated by the rank $k$ random tensor model through realizing that the Feynman graphs of a rank $k$ tensor model possess a graphical representation in terms of $(k+1)$-colored graphs, where a new color is associated with the propagator. This provides the desired geometric interpretation of the Feynman graphs of an uncolored rank $k$ tensor model with tessellations $\Delta$ of piecewise linear pseudo-manifolds of dimension $k$. It follows that the partition function of these random tensor models possesses a geometric interpretation
\begin{equation}
    Z=\sum_{\Delta}\,\mathcal A(\Delta)=\sum_{\Delta}\,e^{-(-\ln(\mathcal{A}(\Delta)))},
\end{equation}
where $\mathcal A(\Delta)$ denotes the Feynman amplitude associated with the Feynman graph dual to $\Delta$. This resembles the random lattice partition function for Euclidean quantum gravity
\begin{equation}
    Z_{grav.}=\lim_{a\to 0}\sum_{\Delta} \exp(-S_E[\Delta,a]),
\end{equation}
when the gravity action $S_E[\Delta,a]$ is identified with $-\ln(\mathcal{A}(\Delta))$. The Feynman amplitude depends on the details of the random tensor model, but one can generally say that they depend on the number of $N_k$ of $k$ simplices and the number $N_{k-2}$ of $k-2$ simplices in $\Delta$ as well as the tensor size $N$ and the coupling constants $\lambda_i$. An example amplitude for $k=3$ with one coupling is
\begin{equation}
    \mathcal A(\Delta)=N^{N_1-3N_3/2}\,(\lambda\bar\lambda)^{N_3/2},
\end{equation}
where the couplings $N,\lambda$ possess a simple relation with the couplings in the Regge-expression of General Relativity in three dimensions $S_R[\Delta]=\kappa_3 N_3-\kappa_1 N_1$. Hence, $\kappa_1=\ln(N)$ and $\kappa_3=\frac{3}{2}\ln(N)-\frac 1 2 \ln(\lambda\bar\lambda)$ establishes a relation with the discrete General Relativity coupling constants. 

The total volume is $\langle V\rangle = \langle N_d\rangle\, a^d \, V_o$, where $V_o$ is the filling factor of the geometric building blocks. Hence, one can take the lattice continuum limit $a\to 0$ at fixed total volume $\langle V \rangle$ by tuning to a point where the expectation value of the total volume diverges. This requires that $\langle N_d\rangle$ diverges. It turns out that this in turn requires that $N\to\infty$. However, to obtain simultaneously a finite value of the total volume and of Newton's constant, one needs to tune $\lambda$ and $N$ simultaneously. Since $Z$ diverges for $N\to\infty$ one can only obtain a finite result when $\lambda$ approaches a critical point $\lambda_*$ as $N\to\infty$ is approached. Hence, we can write down the required behaviour of $\lambda(N)=\lambda_*+c\,N^{-\theta}$, where $c$ is an arbitrary constant and $\theta$ the critical exponent. In other words, the conjecture is that the continuum limit of lattice quantum gravity can be investigated by studying the critical points in the large $N$ behaviour of random tensor models.

So far we have only considered a canonical quadratic term $T_{i_1i_2...i_k}\bar T_{i_1i_2...i_k}$, as is implied by $U^k(N)$ invariance. This kinetic term leads to an index-independent propagator $\propto \delta^{i_1j_1}...\delta^{i_kj_k}$, so each closed loop of indices will contribute with a factor of $N$ to the amplitude, but can not depend on the number of vertices that are crossed when going around this loop. However, we will see shortly that such a dependence of the amplitude can be motivated geometrically. To construct tensor models whose amplitude depends non-trivially on the number of vertices crossed by a closed index loop. Before motivating these so-called ``dually weighted" tensor models, we will consider the general setup of the FRGE for tensor models.
\subsection{Application of the FRGE to tensor models}
One of the most convenient tools to investigate critical behaviour is the functional renormalization group equation (FRGE). In the usual setting the FRGE 
\begin{equation}
    \partial \Gamma_k[\phi]=\frac 1 2 \textrm{Tr}\left(\frac{\dot R_k}{\Gamma_k^{(2)}[\phi]+R_k}\right)
\end{equation}
describes how the effective average action
\begin{equation}
    \Gamma_k[\phi]:=\sup_J\left\{J.\phi-\ln(Z_k[J])\right\}-\frac 1 2 \phi.R_k.\phi
\end{equation}
changes when the IR suppression scale $k$ is changed. This IR-suppression scale is introduced through a modification of the bare action by the scale dependent mass term $\Delta_k\,S[\phi]=\frac 1 2 \phi.R_k.\phi$, which is designed to give a mass of $\mathcal O (k)$ to modes in the IR of the scale $k$ while not significantly affecting modes in the UV of this scale. Heuristically, one can argue as follows: $\Delta_k\,S[\phi]$ dominates the path integral in the limit $k\to\infty$ and hence the saddle point approximation of the path integral becomes exact in this limit and shows that the effective average action coincides with the bare action when $k\to\infty$. Hence, one finds critical points as UV fixed points of the FRGE and can study the critical behaviour by studying the linearized flow near the fixed points. 

The usual FRGE arguments outlined above relies heavily on the mass dimension and scaling with a scale $k$ that possesses units of mass. Such a mass scale is missing in the random tensor setup, instead one wants to study the scaling of the couplings with the dimensionless tensor size $N$. This requires one to identify (1) the scaling of the IR-suppression term with $N$ and (2) the scaling of the coupling constants with $N$. It turns out that the requirement that the RHS of the FRGE admits a $1/N$-expansion imposes significant restrictions on the scaling with $N$, but it does not fix it completely. To obtain a completely determined scaling with $N$ one needs to impose that the bare action possesses a geometric interpretation. Essentially, the requirements are that (1) the bare propagator and the modified propagator (after including the IR-suppression term $\Delta_N S[T]$) possess the same scaling for large index values and (2) that the interaction term possesses the scaling necessary for the geometric interpretation. These two initial conditions, together with the restrictions that result from the $1/N$-expandability of the RHS of the FRGE fix the setting that is sufficient to investigate the large-$N$-critical behaviour.   

\subsection{2D Causal Matrix Model}
A matrix model that enforces a preferred time slicing in its Feynman-graphs was proposed by Benedetti and Henson in \cite{Benedetti:2008}. This model is constructed using two dynamical $N\times N$ matrices, $A$ and $B$, representing the spacelike and timelike edges of a triangle, and a constant matrix $C$ which implements the dual weighting. The partition function is    
    \begin{equation}
        Z=\int dAdB \, e^{-NTr\left(\frac{1}{2}A^2+\frac{1}{2}(C^{-1}B)^2-gA^2B\right)}\label{cdtmmoriginal},
    \end{equation} 
where in the large $N$ limit the matrix $C$ must satisfy the condition   
    \begin{equation}\label{trC2}
        Tr(C^m)=N\delta_{2,m},
    \end{equation} 
with $m\in\mathbb{N}$. The partition function  \eqref{cdtmmoriginal} with a weighting matrix $C$ that implements \eqref{trC2} generates Feynman diagrams that possess the geometric interpretation of polytopes with an arbitrary number of space-like edges and only two time-like edges (see fig. 1). This is clear by analyzing the free propagators $(g=0)$  of the model
 \begin{equation}
    \langle A_{ij}A_{kl}\rangle_0=\frac{1}{N}\delta_{il}\delta_{kj},
\end{equation}
\begin{equation}
    \langle B_{ij}B_{kl}\rangle_0=\frac{1}{N}C_{il}C_{kj},
\end{equation}
\begin{equation}
    \langle A_{ij}B_{kl}\rangle_0=0.
\end{equation}
As we can see in fig. 1, this restriction implements a foliation of the discrete geometries that appear in the expansion of the partition function, thus introducing the structure necessary for the implementation CDT in tensor models.
\begin{figure}[H]
  \centering
    \includegraphics[width=0.45\textwidth]{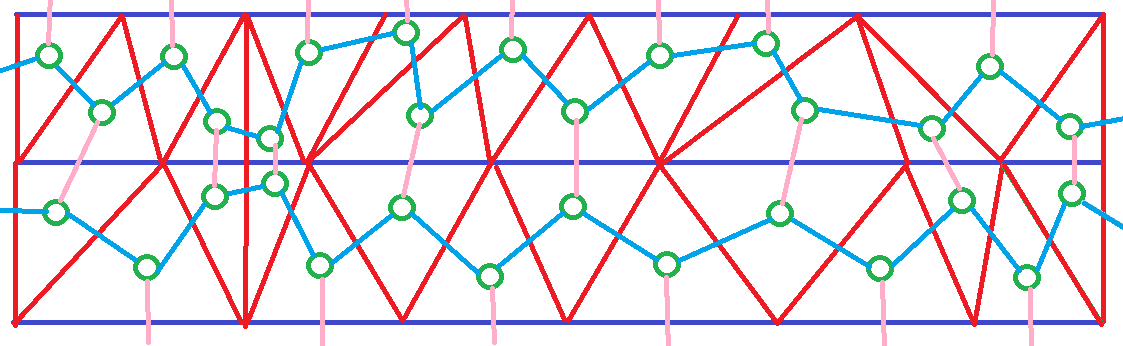}
  \label{fig1dualtopchange}
  \caption{Part of a dual triangulation to a Feynman graph. The solid colors red and blue indicate the time- and space-like boundaries of dual triangles, the light colors the dual propagators in the Feynman graph and the green circles the vertices in the Feynman graph. The important fact to note is that having precisely two pink propagators in each closed loop implies that the blue lines foliate the entire Feynman graph. Notice that we drew the propagators as single lines to not clutter the picture too much; the usual depiction of the matrix model propagator would be though a double line, i.e. line one for each contacted index.}
\end{figure}
We proceed by integrating out matrix $B$ since it is a gaussian integral, obtaining   
    \begin{equation}
        Z=\int dA\, e^{-NTr\left(\frac{1}{2}A^2-\frac{g^2}{2}(A^2C)^2\right)}.
    \end{equation}
This partition function together with condition \eqref{trC2} determines our starting point in this work.\par
One can understand the integration over matrix $B$ as the gluing of triangles along their spacelike edges. This gives rise to a model of squares only with timelike edges. This produces an an-isotropic quadrangulation with rigidity associated with condition \eqref{trC2}.    
\begin{figure}[H]
  \centering
    \includegraphics[width=0.3\textwidth]{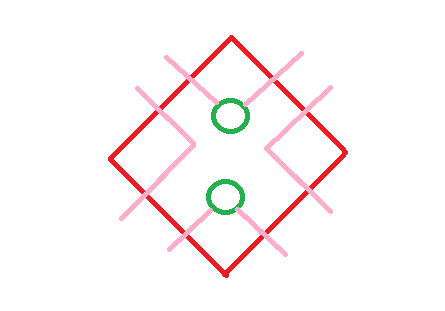}
  \label{quad}
  \caption{Resulting vertex after the integration over matrix $B$. The double pink lines indicate the propagator and the green circles indicate the insertion of the matrix $C$ in the interaction.}
\end{figure} 
We identify the matrix model action that implements CDT as
\begin{equation}
S=\frac{1}{2}Tr(AA^\top)-\frac{g^2}{2}Tr(AA^\top CAA^\top C), \label{CDTaction}
\end{equation}
which takes the form of the Euclidean action \cite{DiFrancesco:1993cyw}, except for the presence of matrix $C$. The appearance of the dual weighting matrix $C$ changes the symmetry of the matrix model. Let us consider an $N\times N$ orthogonal matrix, $O$, and the transformation
\begin{equation}
\begin{split}
A & \rightarrow AO \\
A^\top & \rightarrow O^\top A^\top \label{symmetry}.
\end{split}
\end{equation} 
Since the combination $AA^\top$ is invariant, \eqref{symmetry} is a symmetry of the Euclidean and the CDT action, however the conjugate symmetry under $A \to O\,A$ and $A^\top \to A^\top\,O^\top$ is {\it only} a symmetry of the Euclidean action and {\it not} of the CDT action. This shows an explicit difference with the real Euclidean matrix model. In the language of the renormalization group: the CDT action \eqref{CDTaction} lives in a different theory space, which is governed by a different symmetry.
\subsubsection{Weighting matrix}

The matrix $C$ implements the weighting of closed loops of propagators in the Feynman graph expansion, i.e. the dual weighting of the Feynman graphs. In principle one could define this matrix abstractly only through the property \eqref{trC2} and only use equation \eqref{trC2} whenever the matrix occurs in a Feynman diagram. However, in order to do practical calculations with the FRGE, it is very useful to have an explicit representation of $C$ at ones disposal.  

For a $N\times N$ diagonal matrix, $X$, with eigenvalues $\lbrace x_i\rbrace$ we can write its characteristic polynomial as
\begin{equation}
    P_X(t)=\sum_{k=0}^N(-1)^{k}e_{k}t^{N-k},
\end{equation}
where $e_k$ is
\begin{equation*}
\begin{split}
    e_0(x_1,...,x_N) & = 1,\\
    e_1(x_1,...,x_N) & = \sum_{i=1}^N x_i,\\
    e_2(x_1,...,x_N) & = \sum_{1\leq i\leq j\leq N} x_ix_j,\\
    \vdots \\
    e_N(x_1,...,x_N) & = x_1x_2... x_N,
\end{split}
\end{equation*} then Newton identities allow us to write this coefficients in terms of the $k$-th power of the trace of $X$, $p_k$, in the form  
\begin{equation}
 ke_k=\sum_{i=1}^k(-1)^{i-1}p_ie_{k-i},   
\end{equation}
so, $C$ can be found by solving 
\begin{equation}
    P_C(t)=t^N-\frac{1}{2}Nt^{N-2}+\frac{1}{8}N^2t^{N-4}+...=0.
\end{equation}
These solutions exist by the fundamental theorem of algebra and one can use ones preferred approximation scheme to obtain these. One scheme that suggests itself in particular when one wants to gain insights into the effects of dual weightings is to build a matrix $C$ from smaller blocks of matrices $C_o$, so $C=diag(C_o,...,C_o)$. The matrix obtained in this way does not implement the entire tower of equations \eqref{trC2}, but permits traces periodically. This approach is particularly interesting to study, since it allows us to study how many of the equations one needs to enforce to attain the phase transition between the Euclidean matrix model and the CDT matrix model.
The first three matrices $C_o$ are 

\begin{itemize}

\item $k=2$:
\begin{equation}
    C_o=diag(-1,1),\label{CN=2}
\end{equation}
    \item $k=4$:
\begin{equation}
\begin{split}
C_o=diag(-1\ldotp09-0\ldotp45i, -1\ldotp09+0\ldotp45i, \\
1\ldotp09-0\ldotp45i, 1\ldotp09+0\ldotp45i),\label{CN=4}
\end{split}
\end{equation}
\item $k=6$:
\begin{equation}
\begin{split}
    C_o=diag(1\ldotp02+0\ldotp70i, 1\ldotp02-0\ldotp70i, 1\ldotp37,\\ -1\ldotp02+0\ldotp70i, -1\ldotp02-0\ldotp70i, -1\ldotp37).\label{CN=6}
\end{split}
\end{equation} 
\end{itemize}
Putting these together as blocks to build an $N \times N$ matrix gives for example for $k=2$
   \begin{equation}
    C_{jj}=(-1)^{j},\label{c2}
\end{equation} and for $k=4$
\begin{equation}
C_{jj}=
\left \{
\begin{aligned}
(-1)^k\gamma-i\xi, \quad j & = 2k-1,\\
(-1)^k\gamma+i\xi, \quad j & = 2k,\label{c4}
\end{aligned}
\right.
\end{equation} that are $N \times N$ matrices formed by $2\times 2$ and $4 \times 4$ blocks, and where $\gamma=1.02$ and $\xi=0.70$.

\subsection{Functional Renormalization Group for Matrix Models}

Let us briefly review the application of the FRGE to matrix and tensor models. One can follow the fundamental presentation of \cite{Wetterich:1992} and apply it to matrix models as done in \cite{Eichhorn:2013isa}. The starting point is the definition of the effective average action $\Gamma_N[\phi]$ in the presence of an IR-suppression term $\Delta S_N[\phi]$:
\begin{equation}
    \Gamma_N[\varphi]=inf_J\{ W_N[J]+J\varphi-\Delta S_N[\varphi]\},\label{EAvAction}
\end{equation} where
\begin{equation}
    \exp{\left(-W_{N}[J]\right)}=\frac{1}{\mathcal{N}_{N}}\int_\Lambda\mathcal{D}\phi e^{-S[\phi]+J\phi-\Delta S_N[\phi]},
\end{equation} 
where $\varphi$ represents the expectation value of a quantum field $\phi$, while the term
\begin{equation}
    \Delta S_N[\phi]=\frac{1}{2}\phi_{ab}R_N^{abcd}\phi_{cd}
\end{equation} 
represents an IR-suppression term in so far as it is designed to give a mass term of order $N$ to ``IR" degrees of freedom of the matrix. Since matrix and tensor models do not implement a fundamental scale, there is no canonical identification of which degrees of freedom are ``IR". Rather one needs to implement by hand a division of theory space according to an RG scale $k$. The simplest assignment is to identify the upper-left corner of the matrix with index values below the scale $k$ as ``IR". Once the IR-suppression term is chosen, one can proceed as in \cite{Wetterich:1992}; one arrives at the FRGE
\begin{equation}
\partial_t{\Gamma}_N=\frac{1}{2}Tr\left(\frac{\partial_tR_N}{R_N+\Gamma_N^{(2)}}\right),\label{WetterichEq}
\end{equation} 
where $t=\ln N$. The solutions to \eqref{WetterichEq} are functionals of the $N\times N$ matrix $\phi$ and hence of infinitely many degrees of freedom in the large $N$-limit. Practically one resorts to finite truncations of the effective average action, i.e. one performs an expansion of the effective average action into monomials   
\begin{equation}
    \Gamma_k[\varphi]=\sum_i \bar{g}_i(k) \mathcal{O}_{i}[\varphi].\label{anstrunc}
\end{equation} 
Then one truncates this expansion at a manageable set of operators $\mathcal{O}_{i}$. In this way one reduces he study of the flow to a projected flow in the space of coupling constants $\bar g_i$. The quality of the FRGE results depends critically on the operators that are included in the truncation. In matrix models it turned out that surprisingly good approximations to the FRGE flow where obtained in \cite{Eichhorn:2013isa} by considering the flow of single trace operators. The analogous truncation in the presence of the matrix $C$ is
\begin{equation}
\Gamma_N=\frac{Z}{2}Tr(AA^\top)+\sum_{n=2}^{P}\frac{\bar{g}_{2n}}{2n}Tr((AA^\top C)^2(AA^\top)^{n-2})\label{gammac2P},
\end{equation} 
which only includes operators with $AA^\top$, which is invariant under \eqref{symmetry}, and two $C$ matrices. In the present contribution we will truncate this to the ansatz that contains the bare action and the single trace operator that can directly contribute to the beta functions of the bare action at one loop. This truncation is:
\begin{equation}
\begin{split}
    \Gamma_N =\frac{Z}{2}Tr(AA^\top)+\frac{\bar{g}_4}{4}Tr(AA^\top CAA^\top C)\\
    +
     \frac{\bar{g}_6}{6}Tr(AA^\top CAA^\top CAA^\top)\label{gammac2}.
     \end{split}
\end{equation} 
We introduce the dimensionless couplings
\begin{equation}
\begin{split}
\bar{g}_4 & = Z^2N^{\alpha_4} g_4 \\
\bar{g}_6 & =Z^3N^{\alpha_6} g_6 \label{reescaling}
\end{split}
\end{equation} 
where $\alpha_4$ and $\alpha_6$ are as of yet undetermined, since  the matrix model does not include an intrinsic notion of scale. The scale is later fixed by imposing that the beta functions admit a $1/N$ expansion. 

To make the calculation concrete, we choose the explicit form of the IR-suppression term  $R_N$ to take the form
\begin{equation}
R^{abcd}_N=Z\left(\frac{N}{a+b}-1\right)\theta\left(1-\frac{a+b}{N}\right)\delta^{ac}\delta^{bd},\label{LitimR}
\end{equation} 
which has the advantage of being a diagonal and field independent tensor, so we can readily invert the kinetic term to obtain the propagator. It is practically useful to split the second variation of the effective average action into a field independent term $G$ and a field dependent term $F$:
\begin{equation}
R_N+\Gamma^{(2)}_N=G_N+\bar{g}_4F^{(4)}_N[A]+\bar{g}_6F^{(6)}_N[A]\label{separaciongamma},
\end{equation} 
which allows us to expand the RHS of the Wetterich equation as a geometric series, using only the propagator $P=G^{-1}$ and the $F$-term:
\begin{equation}
\begin{split}
\frac{1}{2}Tr\left(\frac{\dot{R}_N}{R_N+\Gamma_N^{(2)}}\right)=\frac{1}{2}\sum_{k=0}\left((-1)^kTr(\dot{R}P(FP)^k)\right)\\=\frac{1}{2}Tr(\dot{R}P)-\frac{1}{2}Tr(\dot{R}PFP)+\frac{1}{2}Tr(\dot{R}PFPFP)\\-\frac{1}{2}Tr(\dot{R}PFPFPFP)+...\label{splitW}
\end{split}
\end{equation} 
The upshot of this $P-F$ expansion is that each $F$ term contributes more field operators. Hence a truncation that contains polynomial operators with only a finite number of fields terminates the geometric series at a finite number of terms. With the proposed truncation \eqref{gammac2}, the first term in \eqref{splitW} is of order zero in the Feynman diagrams expansion since there is no field contribution, the second term gives rise to 2-vertex and 4-vertex diagrams which contribute to $\eta$ and $\beta_4$, the third one to 4-vertex and 6-vertex diagrams which contribute to $\beta_4$ and $\beta_6$ and so on.

\subsection{Benchmark results}\label{exactcritexp}

We use the FRGE to find fixed points of the RG flow and to investigate the universality class associated with this fixed point. This is done by calculating the critical exponents $\theta$ at the fixed point, i.e. by considering the linearized FRGE-flow at the fixed point, where the critical exponents appear as the eigenvalues of the Hessian of the beta functions. We chose our truncation in such a way that we can resolve the fixed point that known as the double scaling limit in the matrix model literature. This fixed point possesses a single positive critical exponent, which is usually expressed in terms of the string susceptibility $\gamma_{str}$:
\begin{equation}
\theta=\frac{2}{2-\gamma_{str}}.
\end{equation}
For Euclidean Matrix Models \cite{DiFrancesco:1993cyw}  $\gamma_{str}=-\frac{1}{2}$, while for CDT \cite{Ambjorn:1998} $\gamma_{str}=+\frac{1}{2}$, which leads to the following critical exponents
\begin{equation}
\theta_{MM}=\frac{4}{5}, \qquad \theta_{CDT}=\frac{4}{3}.  \label{theocritexp} 
\end{equation}

\section{$\beta$-functions}\label{sec:betaFunctions}

In this section we summarize the steps that we took to obtain the beta functions of the matrix model for CDT.

\subsection{Operator products}\label{fterms}

The structure of the beta functions is determined by the operator products of the $F$-terms that we showed  in \eqref{splitW}. The terms $F^{(4)}_N$ and $F^{(6)}_N$ are the second variations of the operators whose contribution to the effective average action is measured by the coupling constants $\bar{g}_4$ and $\bar{g}_6$. The second variations take the form

\begin{equation}
F^{(4)}_N=C^{ac}(A^\top CA)^{db}+(CA)^{ad}(CA)^{cb}+(CA{A^\top}C)^{ac}\delta^{db}\label{F4CAUS}
\end{equation}
and
\begin{equation}
\begin{split}
F^{(6)}_N=\delta^{ac}({A^\top}CA{A^\top}CA)^{db}+A^{ad}(CA{A^\top}CA)^{cb}\\
+(A{A^\top}C)^{ac}({A^\top}CA)^{db}+(A{A^\top}CA)^{ad}(CA)^{cb}\\
+(A{A^\top}CA{A^\top}C)^{ac}\delta^{db}+C^{ac}({A^\top}CA{A^\top}A)^{db}\\
+CA)^{ad}(CA{A^\top}A)^{cb}+(CA{A^\top}C)^{ac}({A^\top}A)^{db}\\
+(CA{A^\top}CA)^{ad}A^{cb}+(CA{A^\top}CA{A^\top})^{ac}\delta^{db}\\
+C^{ac}({A^\top}A{A^\top}CA)^{db}+(CA)^{ad}(A{A^\top}CA)^{cb}\\
+(CA{A^\top})^{ac}({A^\top}CA)^{db}+(CA{A^\top}A)^{ad}(CA)^{cb}\\
+(CA{A^\top}A{A^\top}C)^{ac}\delta^{db}\label{F6CAUS}.
\end{split}
\end{equation}
By looking at \eqref{splitW}, we notice that traces of products of \eqref{F4CAUS} and \eqref{F6CAUS} give rise to a big range of operators which are not present in the truncation ansatz \eqref{gammac2}, such as $\left(Tr(CA)\right)^2$, $Tr(CAA^\top C^2)Tr(A^\top CA)$, etc. However, a reasonable projection rule onto the truncation should not project these operators onto the beta functions of the truncation. We therefore analyze which operators can contribute to the beta functions in the truncation, i.e. to $\beta_4$ and $\beta_6$. Considering for example the trace
\begin{equation}
    Tr\left( F^{(4)}_N\right)=Tr(C)Tr(A^\top CA)+Tr(CA{A^\top}C)Tr(\delta),\label{trf4}
\end{equation} 
we see that each term contains the matrix $C$, while we know from the structure of the P-F-expansion that these are the only terms generated by the restriction of the FRGE to the truncation that contain two matrices $A$. Hence, the restriction of the FRGE to the truncation does not generate terms $\sim Tr(AA^\top)$ and hence does not generate any term that contributes to the anomalous dimension $\eta$. To generate a contribution to the anomalous dimension, one needed to include a term with a single $C$ matrix in the truncation. This term would then be generated at one loop by the first term in \eqref{trf4} and in turn contribute to $\eta$ at one loop. The investigation of this kind of secondary effect however goes beyond the scope of this first investigation.

This analysis relies on the fact that our projection rule is able to discern the structure in which the matrices $A$ are contracted with the constant weighting matrix $C$, so at first sight one might worry that such a projection does not exist. However, one can consider the appearance of the matrix $C$ in the operators as a special case of operators with index-dependence, i.e. operators whose variations w.r.t. $A$ can not be expressed in terms of $A$ and $\delta_{ij}$, which can be discerned by a suitable projection rule. Hence it is not only possible, but even prudent to use a projection rule that discerns the different ways in which the matrix $C$ is contracted.

To make this distinction, we mark in the following the terms that contribute to the beta functions in our truncation by putting a box around them. Subsequently, we will impose the use of a projection rule that only retains these operators and thus consider only the contributions of the boxed terms.
    
\begin{equation}
\begin{split}
     Tr\left( F^{(6)}_N\right)=\boxed{Tr(\delta)Tr(AA^\top CAA^\top C)}\\
     +\boxed{3Tr(AA^\top CAA^\top C)Tr(\delta)}\\
     +2Tr(C)Tr(A^\top CAA^\top A),
\end{split}     
\end{equation}
\begin{equation}
\begin{split}
    Tr\left( F^{(4)}_NF^{(4)}_N\right)=\boxed{Tr(CC)Tr(A^\top CAA^\top CA)}\\
    +Tr(CAA^\top CCAA^\top C)Tr(\delta),
\end{split}    
\end{equation}
\begin{equation}
\begin{split}
    Tr\left( F^{(4)}_NF^{(6)}_N\right)=\boxed{2Tr(CC)Tr(A^\top CAA^\top AA^\top CA)}\\
    +3Tr(CAA^\top CAA^\top CAA^\top C)Tr(\delta)\\  +Tr(C)Tr(A^\top CAA^\top CAA^\top CA),
\end{split}    
\end{equation}
\begin{equation}
\begin{split}
    Tr\left( F^{(4)}_NF^{(4)}_NF^{(4)}_N\right)=Tr(C^3)Tr(A^\top CAA^\top CAA^\top CA)\\
    +Tr(CAA^\top CCAA^\top CCAA^\top C)Tr(\delta).
\end{split}
\end{equation}

When considering $Tr\left((F^{(4)}_N)^n\right)$ with $n>2$ we see that all resulting operators contain at least three $C$ matrices which are not present in the original proposed action \eqref{gammac2}, this means that in this truncation the $\beta$-functions do not possess contributions coming from these traces. 

\subsection{General Form of the $\beta$-functions} \label{betaformal}

Now that we have identified the terms that can contribute to the $\beta$-functions, we can write down the general structure of the beta functions. To do so, we introduce the constants $D_i$, $E_i$ and $F_i$, which depend on the details of the projection rule. Repeating the same argument as in the previous subsection for the single trace truncation \ref{gammac2P} we obtain
\begin{equation}
    \eta =0,
\end{equation}
for $i$ odd\begin{equation}
\beta_{2i}=(i\eta-\alpha_{2i})g_{2i}+D_i g_{2(i+1)}+E_i g_{2i}g_{4},
\end{equation}
for $i$ even\begin{equation}
\beta_{2i}=(i\eta-\alpha_{2i})g_{2i}+D_i g_{2(i+1)}+E_i g_{2i}g_{4}+F_i g_{(i+2)}^2.
\end{equation}
We can see in particular that in this truncation tadpoles and 2-vertex diagrams contribute.

\section{Fixed point analysis and scheme dependence}\label{sec:fixedPointAnalysis}

By using the obtained general form of the beta functions for the single trace truncation at our disposal we can discuss fixed points. We first consider the fixed point structure analytically, before inserting particular truncation rules, which provide numerical values for the critical exponents, which allows us to discuss the scheme dependence of our calculation.

\subsection{Analytic fixed point analysis}

By setting our truncation to \eqref{gammac2}, we obtain the following beta functions

\begin{equation}
    \beta_4=(-\alpha_4)g_4+D_2g_6+E_2{g_4}^2,
\end{equation}
\begin{equation}
    \beta_6=(-\alpha_6)g_6+E_3g_4g_6,
\end{equation} where $\eta$ has been set to zero in accordance with our previous analysis. The set of fixed points of this system of beta functions is
\begin{equation}
 \left({g_4}^*,{g_6}^*\right)=\left\lbrace(0,0),\left(\frac{\alpha_4}{E_2},0\right),\left(\frac{\alpha_6}{E_3},\alpha_6\frac{\alpha_4E_3-\alpha_6E_2}{D_2{E_3}^2}\right)\right\rbrace,   
\end{equation} 
and the Hessian matrix (defined as $H_{ij}=-\frac{\partial\beta_i}{\partial g_j}$) is
\begin{equation}
\begin{pmatrix}
\alpha_4-2E_2g_4 & -D_2  \\
-E_3g_6 & \alpha_6E_3g_4  \\
\end{pmatrix}    \label{hessian}.
\end{equation}
Hence, the critical exponents, i.e. the eigenvalues of the Hessian, evaluated in each of the fixed points take the form
\begin{table}[H]
\centering
 \resizebox{8.7cm}{!} {
\begin{tabular}{| c | c |}
\hline
Critical point & Critical exponents \\
\hline \hline
\multirow{2}{3cm}{$(0,0)$} & $\alpha_4$ \\ \cline{2-2}
& $\alpha_6$ \\ \hline
\multirow{3}{3cm}{$\left(\frac{\alpha_4}{E_2},0\right)$} & $-\alpha_4$ \\ 
\cline{2-2}
& \multirow{2}{*}{$\frac{\alpha_6E_2-\alpha_4E_3}{E_2}$}\\
& \\ \hline
\multirow{4}{3cm}{$\left(\frac{\alpha_6}{E_3},\alpha_6\frac{\alpha_4E_3-\alpha_6E_2}{D_2{E_3}^2}\right)$} & \multirow{2}{*}{$\frac{-2\alpha_6E_2+\alpha_4E_3-\sqrt{4\alpha_6({E_2}^2-E_2)-4\alpha_4(E_2E_3-{E_3}^2)+\alpha_4{E_3}^2}}{2E_3}$}\\
& \\
\cline{2-2}
& \multirow{2}{*}{$\frac{-2\alpha_6E_2+\alpha_4E_3+\sqrt{4\alpha_6({E_2}^2-E_2)-4\alpha_4(E_2E_3-{E_3}^2)+\alpha_4{E_3}^2}}{2E_3}$}\\
& \\
\hline
\end{tabular}}
\caption{Critical points with its corresponding pair of critical exponents.}
\label{table:expcriticos}
\end{table}
The canonical dimensions $\alpha_4$ and $\alpha_6$ are not fixed by themselves, but tadpole diagrams show that $\alpha_6$ has to be one dimension of $N$ greater than $\alpha_4$. We identify the Gaussian fixed point $(0,0)$, which will not have a relevant direction. There are two non-Gaussian fixed points: the first fixed point $\left(\frac{\alpha_4}{E_2},0\right)$ contains one relevant and one irrelevant direction if $\frac{E_2}{E_3}>\frac{\alpha_4}{\alpha_6}$, and the third fixed point in Table \ref{table:expcriticos} possesses a relevant and an irrelevant direction if $\frac{E_2}{E_3}<\frac{\alpha_4}{\alpha_6}$. These two points are our candidates for a double scaling limit. Next we will examine them using particular schemes.

\subsection{Scheme dependence}\label{Numresults}

To find concrete critical exponents, we supplement the projection rule with the evaluation of both sides of the FRGE at preferred test matrices $A$. Moreover, we consider the two rigidity matrices obtained by constructing a block diagonal matrix from \eqref{CN=2} and \eqref{CN=4}, namely \eqref{c2} and \eqref{c4} respectively. The specific test matrices that we use for the projection are
\begin{equation}
A^{\delta}_{ab}=\delta_{ab}\theta(N-a),\label{Adelta}
\end{equation}
\begin{equation}
A^{\delta -mod}_{ab}=a\delta_{ab}\theta(N-a),\label{Adeltaab}
\end{equation}
\begin{equation}
    A^{\delta-IR}_{ab}=\delta_{1,a}\delta_{1,b}.\label{AdeltaIR}
\end{equation}
Using three different matrices $A$ allows us to estimate a lower bound for the scheme dependence. We expect this because the three field configurations contain two field configurations with distinct UV behaviour and one manifest IR field configuration. One often assumes that the scheme dependence is an actual approximate measure for the quality of the fixed point analysis, however when comparing with analytic results, we will see that this underestimates the truncation error.

The corresponding obtained critical exponents are shown in Table \ref{res1}, where euclidean values where computed as done in \cite{Eichhorn:2013isa} using \eqref{Adelta}, \eqref{Adeltaab}, \eqref{AdeltaIR} as test fields. 

\end{multicols}

  \begin{table}[H]
    \centering
 {
  \begin{tabular}{ c | c | c | c | c | c | c | c |}
  \cline{2-8} &
    \multicolumn{3}{|c|}{$\delta$} & \multicolumn{2}{|c|}{$\delta$-mod.} & \multicolumn{2}{|c|}{$\delta$-IR} \\
    \hline
    \multicolumn{1}{|c|}{} & Causal$^2$ & Causal$^4$ & Euclidean & Causal$^2$ & Euclidean & Causal$^2$ & Euclidean \\
    \hline
    \multicolumn{1}{|c|}{$\theta$} & 1.033 & 1.008 & 1.046 & 1.024 & 1.033 & 1.052 & 1.065\\
    \multicolumn{1}{|c|}{$\theta'$} & -0.928 & -1.215 & -1.080 & -0.858 & -0.959 & -1.086 & -1.050 \\
    \hline
  \end{tabular}}
  \caption{Numerical values obtained for critical exponents. Causal$^2$ corresponds to values computed using \eqref{c2} and Causal$^4$ corresponds to the ones computed using \eqref{c4}.}\label{res1}
\end{table}

\begin{table}[H]
  \centering
  {
  \begin{tabular}{ c | c | c | c | c | c | c | c |}
  \cline{2-8} &
    \multicolumn{3}{|c|}{$\delta$} & \multicolumn{2}{|c|}{$\delta$-mod.} & \multicolumn{2}{|c|}{$\delta$-IR} \\
    \hline
    \multicolumn{1}{|c|}{} & Causal$^2$ & Causal$^4$ & Euclidean & Causal$^2$ & Euclidean & Causal$^2$ & Euclidean \\
    \hline
    \multicolumn{1}{|c|}{$g^*_4$} & -0.435 & -0.300 & -0.288 & -0.902 & -0.588 & -0.339 & -0.202\\
    \multicolumn{1}{|c|}{$g^*_6$} & -0.118 & -0.094 & -0.061 & -0.387 & -0.208 & -0.040 & -0.026 \\
    \hline
  \end{tabular}}
  \caption{Numerical values obtained for critical points. Causal$^2$ corresponds to values computed using \eqref{c2} and Causal$^4$ corresponds to the ones computed using \eqref{c4}.}\label{table:critpoint}
\end{table}

\begin{multicols}{2}
We see that the relevant critical exponents are all close to $1$, while the irrelevant critical exponents spread a bit wider between $-0.858$ and $-1.215$. Moreover, we observe that all critical exponents lay within the spread obtained by scheme dependence. This means that we can not distinguish the Euclidean models from the "Causal" models built from the $2\times 2$ or $4\times 4$ matrices based on the present derivation of the critical exponents.  

In order to attempt to obtain more accurate numerical values for the critical exponents we use the fixed point approximation. This consists in first finding the zeros of the beta functions, then evaluating the anomalous dimension, $\eta$, in the fixed point $g^*_4$ and substituting this numerical value in the beta functions to find the critical exponents. The critical exponents obtained by using the fixed point approximation are shown in the following table. 
\end{multicols}
  \begin{table}[H]
    \centering
 {
  \begin{tabular}{ c | c | c | c | c | c | c | c |}
  \cline{2-8} &
    \multicolumn{3}{|c|}{$\delta$} & \multicolumn{2}{|c|}{$\delta$-mod.} & \multicolumn{2}{|c|}{$\delta$-IR} \\
    \hline
    \multicolumn{1}{|c|}{} & Causal$^2$ & Causal$^4$ & Euclidean & Causal$^2$ & Euclidean & Causal$^2$ & Euclidean \\
    \hline
    \multicolumn{1}{|c|}{$\theta$} & 0.722 & 0.902 & 0.630 & 0.649 & 0.544 & 0.658 & 0.602\\
    \multicolumn{1}{|c|}{$\theta'$} & -0.953 & -1.222 & -1.116 & -0.880 & -0.995 & -1.105 & -1.090 \\
    \hline
  \end{tabular}}
  \caption{Numerical values obtained for critical exponents using the fixed point approximation. Causal$^2$ corresponds to values computed using \eqref{c2} and Causal$^4$ corresponds to the ones computed using \eqref{c4}.}\label{resmethod2}
\end{table}
\begin{multicols}{2}
Since the numerical values reported in Table \ref{res1} were found to have a strong scheme dependence, it is important to compare the renormalization scheme dependence versus the causal-euclidean results in the latter ones in Table \ref{resmethod2}. We compute the average of the difference between the critical exponents obtained in the different renormalization schemes and the ``Causal vs. Euclidean" results with each of both methods. 
\begin{table}[H]
    \centering
    \begin{tabular}{|c|c|c|}
    \cline{1-3}
           & Renorm. Scheme & Causal vs. Euclidean  \\
           \hline
          full & 0.019  &  0.012 \\
          \hline
          fixed p. a. & 0.049  &  0.084 \\
          \hline
    \end{tabular}
    \caption{Renormalization scheme dependence and Causal-Euclidean difference with both methods.}
    \label{tab:my_label}
\end{table}

We observe that, while the first method (full) shows a stronger renormalization scheme dependence, with the fixed point approximation method the ``Causal vs. Euclidean" relation is more significant than the renormalization scheme dependence.  Regarding the accuracy of the values for the critical exponent obtained with both methods compared to the theoretical values \eqref{theocritexp}, we observe that the Causal ones differ more from the theoretical value than the Euclidean critical exponents. Therefore we can conclude that in this case the fixed point approximation is more useful for differentiating the Causal from the Euclidean results, while the full method reproduces more accurate numerical results. 

\section{Conclusions}\label{sec:conclusions}

This contribution is motivated by the observation that the application of the FRGE to tensor models with dual weights could lead to an approach to quantum gravity that combines the advantages of the systematic search of continuum limits with the FRGE with the physically promising phase diagrams of CDT and EDT. The systematic development of these tools and the systematic investigation of these models is a very ambitious task. In this contribution we took a first step into this direction and considered the FRGE flow of a matrix model for CDT in 1+1 dimensions proposed by Benedetti and Henson. This model implements a foliation through a dual weighting of Feynman graphs, which introduces an action with an index-dependent propagator, which, upon integration of an auxiliary field, introduces an action with an index dependent interaction. 

Recalling the critical exponents analysis in Section 4.1 and the values in Table \ref{table:expcriticos}, the model uses a rigidity matrix $C$ which is chosen in such a way that one of the conditions for the beta function polynomials $\frac{E_2}{E_3}>\frac{\alpha_4}{\alpha_6}$ or $\frac{E_2}{E_3}<\frac{\alpha_4}{\alpha_6}$ is satisfied. In this case we obtain a relevant and an irrelevant directions simultaneously, implying that the only Feynman diagrams that contribute to the partition function are the ones where all $C$ matrices are contracted as $Tr(C^2)$, which is precisely the condition that implements a foliation. In this contribution we considered a single trace truncation in which we included operators that contain two $C$-matrices with the pattern prescribed by the interaction in the Benedetti-Henson model and calculated the beta functions for this truncation. We found that wave function renormalization does not occur in this truncation, since the one-loop structure of the FRGE can only remove one $C$ matrix. This technical observation has far reaching consequences for the structure of the beta functions, which change significantly compared to the Euclidean model, which is obtained by setting $C$ to the identity matrix. We then investigated this system of beta functions in a truncation in which we included only a four- and a six-point interaction. Despite the significant difference in the structure of the beta functions, we found that this truncation contains fixed points that possess the properties of the double scaling limit, which we investigated numerically using three distinct field configurations for projection. 

This numerical investigation revealed a practical challenge: To obtain numerical values for the beta functions one can not resort to an abstract definition of the rigidity matrix $C$, but one requires an explicit numerical expression. We took this as an opportunity to investigate the weakening of the condition $Tr(C^m)=\delta_{m\,\textrm{ mod }k,2}$ for $k=2,4,6$. This has the implication that not all Feynman diagrams without a foliation structure are suppressed, but only a part of these. In particular the case $k=2$ does not introduce any new restriction at the level of Feynman diagrams of the Euclidean model, however, since we used the structure of the beta function for general $C$, we still obtained equations that differ from the Euclidean matrix model. The numerical investigations however revealed that we can not discern the Euclidean and the CDT model on the basis of the critical exponents at the fixed point associated with the double scaling limit. These results are summarized in table 2. We see that the relevant critical exponent $(\theta)$ in the CDT model differs from the the exact values in \ref{exactcritexp} by 0.29 and 0.24 from the Euclidean one. We also observe a $4\%$ spread of these values depending on the field configuration used for projection, which is a significantly lower spread than the difference with the analytic values. This is however consistent with the results obtained in \cite{Eichhorn:2013isa}, where a similar difference from the analytic values was found.

We interpret our results as a encouragement for the investigation of dually weighted tensor models for quantum gravity: Already with the rather simple and elementary techniques used in this contribution we were able to investigate qualitatively the double scaling limit of the dually weighted matrix model; analogous dually weighted tensor models such as the one proposed in \cite{Benedetti:2011nn} can thus be treated with the FRGE in a similar fashion. Our results indicate some practical advise for these future investigations:
\begin{enumerate}
    \item The one-loop structure of the FRGE can only couple effective operators that differ by an index-dependent contraction between two adjacent tensors, not more. Therefore, to study the influence of an operator with index-dependence in more than one contraction one needs to include sufficient "intermediary" operators in the truncation.
    \item For analytical investigations it is possible to work with abstract rigidity structures, that are defined through its properties, such as $Tr(C^m)=\delta_{m,2}$, however for numerical investigations one needs to find (or at least approximate) numerical representations of this abstract structure. One might thus prefer the investigations of models for which one has one of these numerical representations at ones disposal.
    \item If our present observations about the double scaling limit are transferable then one sees that the existence of a fixed point with certain characteristics can be found in rather small truncations. However, the critical exponents found in these truncations can be expected to differ significantly from the exact values (which of course should be accessible through lager truncations and optimized renormalization schemes).
\end{enumerate}


\begin{thebibliography}{xxx}
\bibitem{Benedetti:2008}
  D.~Benedetti and J.~Henson,
  ``Imposing causality on a matrix model,''
  Phys.\ Lett.\ B {\bf 678} (2009) 222
  doi:10.1016/j.physletb.2009.06.027
  [arXiv:0812.4261 [hep-th]].
  \bibitem{Hawking:1979ig}
  S.~W.~Hawking and W.~Israel,
  ``General Relativity : An Einstein Centenary Survey,'' 1979
  \bibitem{Donoghue:1994dn}
  J.~F.~Donoghue,
  ``General relativity as an effective field theory: The leading quantum corrections,''
  Phys.\ Rev.\ D {\bf 50} (1994) 3874
  doi:10.1103/PhysRevD.50.3874
  [gr-qc/9405057].
  \bibitem{Reuter:1996cp}
  M.~Reuter,
  ``Nonperturbative evolution equation for quantum gravity,''
  Phys.\ Rev.\ D {\bf 57} (1998) 971
  doi:10.1103/PhysRevD.57.971
  [hep-th/9605030].
  \bibitem{Reuter:2012id}
  M.~Reuter and F.~Saueressig,
  ``Quantum Einstein Gravity,''
  New J.\ Phys.\  {\bf 14} (2012) 055022
  doi:10.1088/1367-2630/14/5/055022
  [arXiv:1202.2274 [hep-th]].
  \bibitem{Eichhorn:2018yfc}
  A.~Eichhorn,
  ``An asymptotically safe guide to quantum gravity and matter,''
  Front.\ Astron.\ Space Sci.\  {\bf 5} (2019) 47
  doi:10.3389/fspas.2018.00047
  [arXiv:1810.07615 [hep-th]].
  \bibitem{Thiemann:2007zz}
  T.~Thiemann,
  ``Modern canonical quantum general relativity,''
  gr-qc/0110034.
  \bibitem{Perez:2012wv}
  A.~Perez,
  ``The Spin Foam Approach to Quantum Gravity,''
  Living Rev.\ Rel.\  {\bf 16} (2013) 3
  doi:10.12942/lrr-2013-3
  [arXiv:1205.2019 [gr-qc]].
  \bibitem{Freidel:2005qe}
  L.~Freidel,
  ``Group field theory: An Overview,''
  Int.\ J.\ Theor.\ Phys.\  {\bf 44} (2005) 1769
  doi:10.1007/s10773-
  \bibitem{Rivasseau:2012yp}
  V.~Rivasseau,
  ``The Tensor Track: an Update,''
  arXiv:1209.5284 [hep-th].
  \bibitem{Rivasseau:2013uca}
  V.~Rivasseau,
  ``The Tensor Track, III,''
  Fortsch.\ Phys.\  {\bf 62} (2014) 81
  doi:10.1002/prop.201300032
  [arXiv:1311.1461 [hep-th]].
  \bibitem{Rivasseau:2016zco}
  V.~Rivasseau,
  ``Random Tensors and Quantum Gravity,''
  SIGMA {\bf 12} (2016) 069
  doi:10.3842/SIGMA.2016.069
  [arXiv:1603.07278 [math-ph]].
  \bibitem{Rivasseau:2011hm}
  V.~Rivasseau,
  ``Quantum Gravity and Renormalization: The Tensor Track,''
  AIP Conf.\ Proc.\  {\bf 1444} (2012) no.1,  18
  doi:10.1063/1.4715396
  [arXiv:1112.5104 [hep-th]].
  \bibitem{Barcelo:2005fc}
  C.~Barcelo, S.~Liberati and M.~Visser,
  ``Analogue gravity,''
  Living Rev.\ Rel.\  {\bf 8} (2005) 12
   [Living Rev.\ Rel.\  {\bf 14} (2011) 3]
  doi:10.12942/lrr-2005-12
  [gr-qc/0505065].
  \bibitem{Witten:1998qj}
  E.~Witten,
  ``Anti-de Sitter space and holography,''
  Adv.\ Theor.\ Math.\ Phys.\  {\bf 2} (1998) 253
  doi:10.4310/ATMP.1998.v2.n2.a2
  [hep-th/9802150].
  \bibitem{Polchinski:1998rq}
  J.~Polchinski,
  ``String theory. Vol. 1: An introduction to the bosonic string,''
  doi:10.1017/CBO9780511816079
  \bibitem{Regge:1961px}
  T.~Regge,
  ``General Relativity Without Coordinates,''
  Nuovo Cim.\  {\bf 19} (1961) 558.
  doi:10.1007/BF02733251
  \bibitem{Wetterich:1992}
  C.~Wetterich,
  ``Exact evolution equation for the effective potential,''
  Phys.\ Lett.\ B {\bf 301} (1993) 90
  doi:10.1016/0370-2693(93)90726-X
  [arXiv:1710.05815 [hep-th]].
  \bibitem{DiFrancesco:1993cyw}
  P.~Di Francesco, P.~H.~Ginsparg and J.~Zinn-Justin,
  ``2-D Gravity and random matrices,''
  Phys.\ Rept.\  {\bf 254} (1995) 1
  doi:10.1016/0370-1573(94)00084-G
  [hep-th/9306153].
  \bibitem{Ambjorn:1998}
  J.~Ambjorn and R.~Loll,
  ``Nonperturbative Lorentzian quantum gravity, causality and topology change,''
  Nucl.\ Phys.\ B {\bf 536} (1998) 407
  doi:10.1016/S0550-3213(98)00692-0
  [hep-th/9805108].
  \bibitem{Eichhorn:2013isa}
  A.~Eichhorn and T.~Koslowski,
  ``Continuum limit in matrix models for quantum gravity from the Functional Renormalization Group,''
  Phys.\ Rev.\ D {\bf 88} (2013) 084016
  doi:10.1103/PhysRevD.88.084016
  [arXiv:1309.1690 [gr-qc]].
  \bibitem{Benedetti:2011nn}
  D.~Benedetti and R.~Gurau,
  ``Phase Transition in Dually Weighted Colored Tensor Models,''
  Nucl.\ Phys.\ B {\bf 855} (2012) 420
  doi:10.1016/j.nuclphysb.2011.10.015
  [arXiv:1108.5389 [hep-th]].
\end{thebibliography}

\end{multicols}

\end{document}